\documentclass[conference]{IEEEtran}
\IEEEoverridecommandlockouts
\usepackage{cite}
\usepackage{amsmath,amssymb,amsfonts}
\usepackage{algorithmic}
\usepackage{graphicx}
\usepackage{textcomp}
\usepackage{xcolor}
\def\BibTeX{{\rm B\kern-.05em{\sc i\kern-.025em b}\kern-.08em
    T\kern-.1667em\lower.7ex\hbox{E}\kern-.125emX}}

\begin{document}

\title {Security of IT/OT Convergence: \\Design and Implementation Challenges}

\author{\IEEEauthorblockN{Bassam Zahran}
\IEEEauthorblockA{\textit{Computer and Information Sciences} \\
\textit{Towson University}\\
Towson, USA \\
bzahran@towson.edu}
\and
\IEEEauthorblockN{Adamu Hussaini}
\IEEEauthorblockA{\textit{Computer and Information Sciences} \\
\textit{Towson University}\\
Towson, USA \\
ahussa7@students.towson.edu}
\and
\IEEEauthorblockN{Aisha Ali-Gombe}
\IEEEauthorblockA{\textit{Computer and Information Sciences} \\
\textit{Towson University}\\
Towson, USA \\
aaligombe@towson.edu}

}

\maketitle

\begin{abstract}

IoT is undoubtedly considered the future of the Internet. Many sectors are moving towards the use of these devices to aid better monitoring, controlling of the surrounding environment, and manufacturing processes. The Industrial Internet of things is a sub-domain of IoT and serves as enablers of the industry. IIoT is providing valuable services to Industrial Control Systems such as logistics, manufacturing, healthcare, industrial surveillance, and others. Although IIoT service-offering to ICS is tempting, it comes with greater risk. ICS systems are protected by isolation and creating an air-gap to separate their network from the outside world. While IIoT by definition is a device that has connection ability. This creates multiple points of entry to a closed system. In this study, we examine the first automated risk assessment system designed specifically to deal with the automated risk assessment and defining potential threats associated with IT/OT convergence based on OCTAVE Allegro-ISO/IEC 27030 Frameworks.

\end{abstract}


\section{Introduction}
Industrial Control Systems (ICS) is a combination of software and hardware designed to execute and manage industrial operations. The Operational Technologies (OT) are the networking devices and protocols serving the ICS internally. The most used technologies in Industrial Control Systems are Supervisory Control and Data Acquisition (SCADA), Programmable Logic Controllers (PLCs), and Plant Distribution Control Systems (DCSs). IIoT is a sub-domain of the Internet of Things (IoT) and they are interconnected devices intended to improve access, productivity, and decision making in ICS systems. The Industrial Internet of Things (IIoT) provides functionalities like observing power consumption, control of leakage, and many others including safety and security. Recently, Information technology and Operational technologies are converging on a larger scale. This creates a union of IT/OT technologies that improve connectivity, data analytics, and real-time information reporting which supports decision making. Since security is considered a major worry with regards to IT/OT convergence, research must continue to utilize better ways to be ready for the progressively intimidating threat. In this study, we are focusing on evaluating a solution to assess vulnerabilities and threats fronting the IT/OT convergence. In this paper, we present the basic functionalities of the newly developed automated risk assessment system for ICS- IIoT systems titled: Industrial Internet of Things Automated Risk Assessment System (IIoT-ARAS). Our tool performs generic information security assessment, vulnerability analysis, and penetration testing based on the Octave Allegro[3] and ISO/IEC 27030[4] (Draft, expected to be published in 2022) risk assessment methodologies. Currently used risk assessment systems are typically built for OT or IoT systems only. To the best of our knowledge, this is the first automated risk assessment system designed specifically to deal with the potential threats associated with IT/OT convergence based on OctaveAllegro-ISO/IEC 27030 methodologies.

\begin{figure*}[ht!]
\centering
\includegraphics[width=1.5\columnwidth,trim=4 4 4 4,clip]{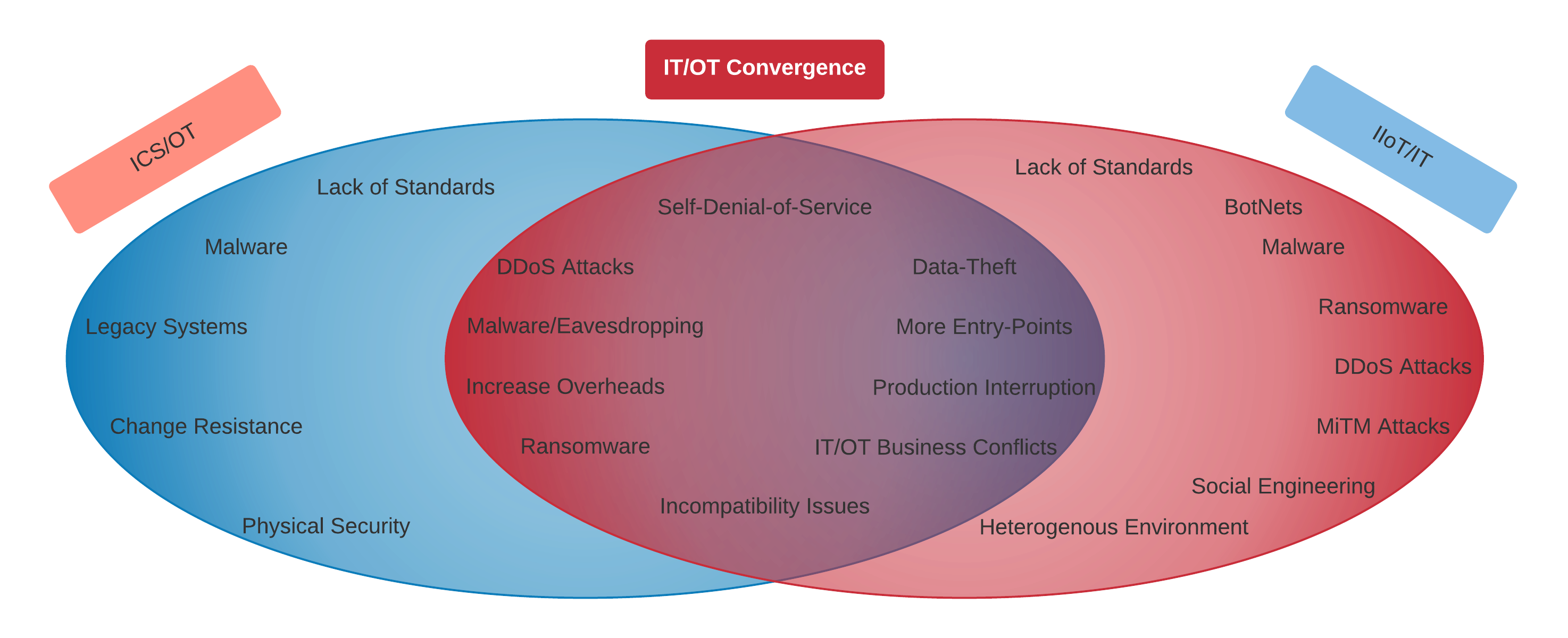}
\caption{Potential Risks in IT/OT Convergence.\cite{b2}}
\label{fig:1}
\end{figure*}

\section{background}

\subsection{ICS Existing Threats and Potential Impacts}
The exceptional nature of ICS systems makes it a puzzling task to modify or suggest improvements to security. Most Industrial Control Systems formats are built to operate for lengthy periods of time with no disruptions. If a malfunction occur, a fail-safe or a reserve system would come in place instantly to ensure continuousness. The main business goal for ICS is to maximizing productivity and eliminating any chance for overheads or delay. This explains the rejection or at least the resistance for any effort to advise changes. 
System steadiness and air-gapping are important requirements in an ICS environment. The air-gapping is isolating the production network from all other external networks. However, with the growing demand for connectivity to the outside world, data collection, cloud computing, and emerging technologies, OT struggles to handle with these changes. It is a known point that risks have a larger impact on ICS/OT than IoT/IT. Over the years, the ICS defensive means depends regularly on being an inaccessible system by creating an air-gap to separate the OT system from the outside world. Besides, investing in physical security. Several studies and researches suggested to ensure redundancy, implementation diversity, and hardening and reinforcing components to avoid tampering\cite{b8}. 
 Currently, this risk dodging technique is not adequate. In fact, even the air-gap setup has become vulnerable to attacks like AirHopper, BitWhisper, GSMem, OOB-CCs, Ramsay, and lately, the Stuxnet\cite{b9}. Figure \ref{fig:1} (the blue section) maps existing challenges/Threats in ICS systems to Impact/Risk-based on the survey of related literature
 \cite{b2}.

\subsection{IIoT Existing Threats and Potential Impacts}
The IIoT gets all vulnerabilities that come with IoT setups. IIoT inserts some more dangers and risks because of the integration with Industrial Control Systems (ICS). While protecting confidentiality, integrity, and availability is the ultimate security goal when bringing devices online, this is not straightforwardly doable when it comes to IIoT.

IIoT networks experience many limitations and constraints that make reaching the CIA model a difficult and complicated task. This is particularly true since the IIoT networks can expand in a large geographical space and be either indoor or outdoor. IIoT security is applied by the vigilant consideration of the confidentiality of information, accuracy, and availability of all objects in the network, security protocols, and capability to connect to objects and devices from a variety of vendors and specifications.

Each of the IIoT layers suffers from several risks. The threats or attacks can be from within the network or from outside and can be through exploiting a vulnerability, misuse, or even human error. In addition to common and developing threats to several of these devices, there is also the possibility of infection through cross-platform malware that might cause to expand the risk to other surrounding networks\cite{b1}.

Efficient defenses and mitigation of possible malware infections and other threats could only be achieved after an in-depth understanding, assessment, and evaluation of the security risk in the IIoT environment. Figure \ref{fig:1} (the red section) lists the most common threats and potential risks in the IIoT network.

\section{IIoT/OT Introduced Threats and Risks}
The idea of security in Industrial Control Systems is based on a risk dodging tactic, where critical systems are isolated from other networks. As a defense mechanism, the OT in ICS is created as an inaccessible system that is air-gaped to assure safe and uninterruptible processes. This technique was adequate and served the need for traditional ICS for a long time without the risk of compromise or security violation except if it was a physical attack on equipment. The key factor of ICSs is to enhance productivity while reducing processing overheads. Unfortunately, security is often considered an overhead in the OT environment. Insertion of IIoT in the industrial arena and the merger between IT and OT has led to an alteration in industrial models. The union of IIoT connectivity and data-oriented techniques into ICS’s process-oriented isolated system has introduced threats into a highly productive ICS network, creating multiple entry points to the supposedly closed environment. Besides, the ICS/OT and IIoT/IT have different business objectivity. Devices in the ICS environment are designed to operate for a lengthy period of time and have a whole domain of legacy equipment still in active use. Systems in the ICS network habitually do not support basic protection methods applied in the implementation of the IIoT systems. For example, authentication and cryptography methodologies are not supported in older, difficult to swap legacy devices and software. Since most ICS devices are special-purpose machines and not general-purpose, it is challenging to introduce customization and implement competent security measures. Installing or updating a security patch to a running ICS system is considered a major task and would not be acknowledged or at least welcomed in such an environment. More importantly, the recent security specifications are more like business practices rather than abiding policies. Therefore, this special nature of IT/OT convergence requires a distinctive and tactical approach to general security and risk assessment.\cite{b2}.
\section{Methodology}
\begin{figure*}[ht!]
\centering
\frame{\includegraphics[width=1.2\columnwidth,trim=0 0 0 2,clip]{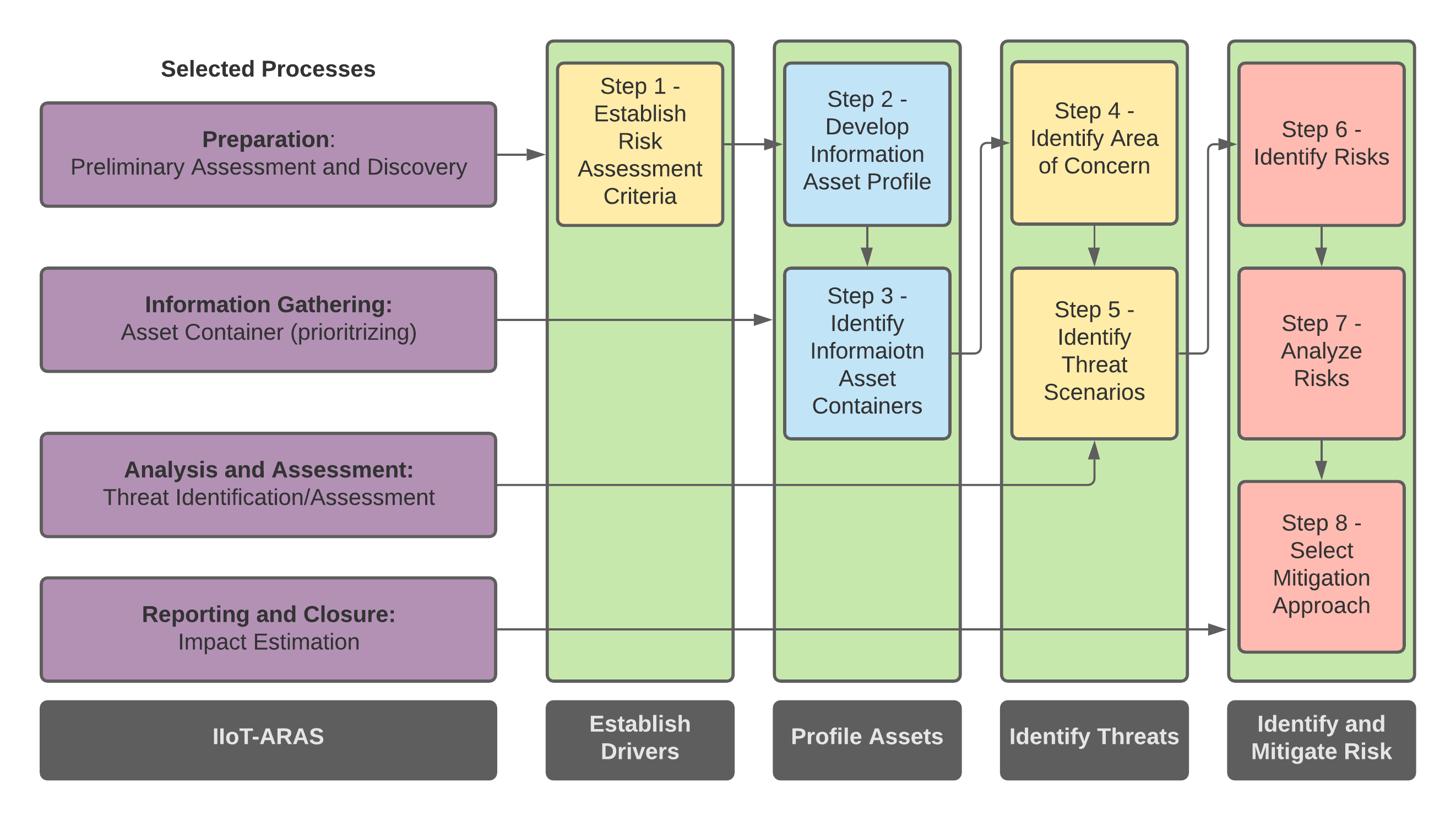}}
\caption{IIoT-ARAS: OCTAVE Allegro Segments}
\label{fig:2}
\end{figure*}
Risk is considered the product of probability of occurrence of an issue and the associated impact on a given entity. A comprehensive comparison shows that each and every existing risk assessment methodology has its own strength and weakness. 
Risk Assessment is crucial in identifying vulnerabilities and potential threats to the system. Risk Assessment methodologies are many, but few tackle the area of IT/OT convergence. Our methodology for IIoT-ARAS is based on customized subset of OCTAVE Allegro and ISO/IEC 27030 frameworks. Combining segments of both guidelines is a trial to circumvent issues related to the heterogeneous IT/OT environment. We have decided to select OCTAVE Allegro and ISO/IEC 27030 since both frameworks complete each other.
\subsection{Risk Assessment Frameworks}
\subsubsection{Octave Allegro}
The OCTAVE Allegro approach is designed to allow a wide range of assessment of operational risk environment to harvest more reliable results without the need for extensive risk assessment knowledge. The approach focuses on information assets in the context of usability, storage areas, data transport, information processing, probability of exposure to threats, vulnerabilities logging, disruptions, and impacts. OCTAVE Allegro consists of guidance, worksheets, and questionnaires. The assessment is done manually by users. Our main focus is to map OCTAVE Allegro assessment processes to be a guideline for an automated risk assessment system.

The OCTAVE Allegro methodology involves eight steps that are organized into four phases, as shown in figure\ref{fig:2}. In phase 1, the risk measurements are defined and mapped to organizational drivers. In the second phase, the prioritization of information assets is based on criticality and importance. The process of profiling assets creates clear boundaries, identifies the security requirements, and recognizes all exact locations where information assets are stored, transported, or processed. In phase 3, based on the previously located information assets,  threats to the information asset are identified. In the final phase, risks to information assets are acknowledged and considered consequently, risk mitigation plans are developed.

OCTAVE Allegro demands that organizations develop asset profiles to enable a more
accurate report of the boundaries of an information asset by ensuring consistency,
clarity, and approved definitions for the asset. In the process of asset profiling,
the company assigns ownership, sets security requirements, and captures the
information asset’s value. A newly created asset profile can be reused, modified, and
updated to match other assets to enhance simplicity and reduce the amount of work for
future assessments. Since the Industrial Internet of Things is communicating with
other devices in the network and with the Industrial Control Systems linked to it
mostly by TCP Stack, asset discovery and profiling is done through a ping sweep,
uPnP, …etc.
The main target of OCTAVE Allegro is to eliminate uncertainty for security requirements. The drawback is that this approach requires human factors to apply solutions to the found risk analysis and identification. 
The current OCTAVE methods exploit threat trees as a guide for identifying threats. While this approach provides a regulated resource for identifying and recording different threat scenarios, users might find them difficult and confusing to use, particularly users with limited risk management experience. The manual process of going through OCTAVE worksheets\cite{b3} to identify vulnerabilities to enable risk identification is very lengthy and can delay considerably the assessment plan.
In practice, users find that performing tool-based vulnerability identification does not provide important additional information that cannot be obtained through scenario identification.\cite{b3}
As a solution, a fully customizable automated risk assessment tool that requires minimum interference from users is highly appreciated.

\subsubsection{ISO/IEC 27030}
ISO/IEC 27030 — Information technology — Security techniques — Guidelines for security and privacy in Internet of Things (IoT)

The standard is being developed to support and guide the assessment of information risk and controls for Internet of Things which in return, partially applicable to Industrial Internet of Things. The standard will be specific to IoT, covering both information security and privacy. IoT devices have the ability to connect to the internet. This might have an impact on the security of the network. Hence, proper security measures and privacy controls are essential. The standard will provide security and privacy guidance for IoT systems, services, and solutions. The standard may also cover device and network trustworthiness and will expectantly align with other IoT standards.\cite{b4}
It is expected that the ISO/IEC 27030 will be comprehensive and covering many areas that are not covered by other frameworks. The standard will be based on ISO/IEC 27005:2011 Series which considered one of the most mature and completed risk assessment methodologies.\cite{b10}
Examples of risks to be addressed by the standard:

\begin{itemize}
  \item High probability of risk impacts, potentially including privacy violation, damage of property, health issues and more;
  \item Many of IoT devices are built to last for a long time. This Long lifecycle of some devices present major issues on ability to secure communication, apply patches, ...etc.;
  \item Shortage in regulating standards with regards to Internet of Things domain, leads to difficulties in managing, monitoring and controlling all devices from different vendors;
  \item Interoperability in a heterogeneous environment and other external networks;
  \item Reduced set of instructions and performance of IoT devices;
  \item Incompatibility among different IoT vendors and manufacturers;
  \item The use of IoT specific device may change with time to adopt with a new scenario. Security measures must adopt with the change.\cite{b4}
\end{itemize}

IoT vendors and manufacturers and to certain extent users, could be unaware of the information risks and required basic controls, hence ISO/IEC 27303 is aiming to raise awareness and push maturity on both the provider and the user sides.
IoT is changing the world in ways that are difficult to predict. This indeed presents a massive privacy threat that has to be controlled to avoid major issues in the future. The standard is due to be published in 2022.

Sensitive and critical sites are susceptible to incidents that would lead to catastrophic impacts. The use of the Industrial Internet of Things to monitor, analyze, and sometimes control industrial systems has an endless number of advantages. On the other hand, ICS systems are used to secure their environment by isolation or air-gapping \cite{b2}. The IIoT creates multiple entry points to a secured close system. IIoT is increasingly used to support and monitor power grids, nuclear plants, the energy sector, and more. IIoT is susceptible to cross-platform malware attacks as well\cite{b1}. These networks have no tolerance for incidents or failure. As the infamous Stuxnet incident confirmed, even isolated, air-gapped critical systems do not completely guarantee their security. 
\begin{figure}
\includegraphics[width=1.0\columnwidth,height=0.37\textheight]{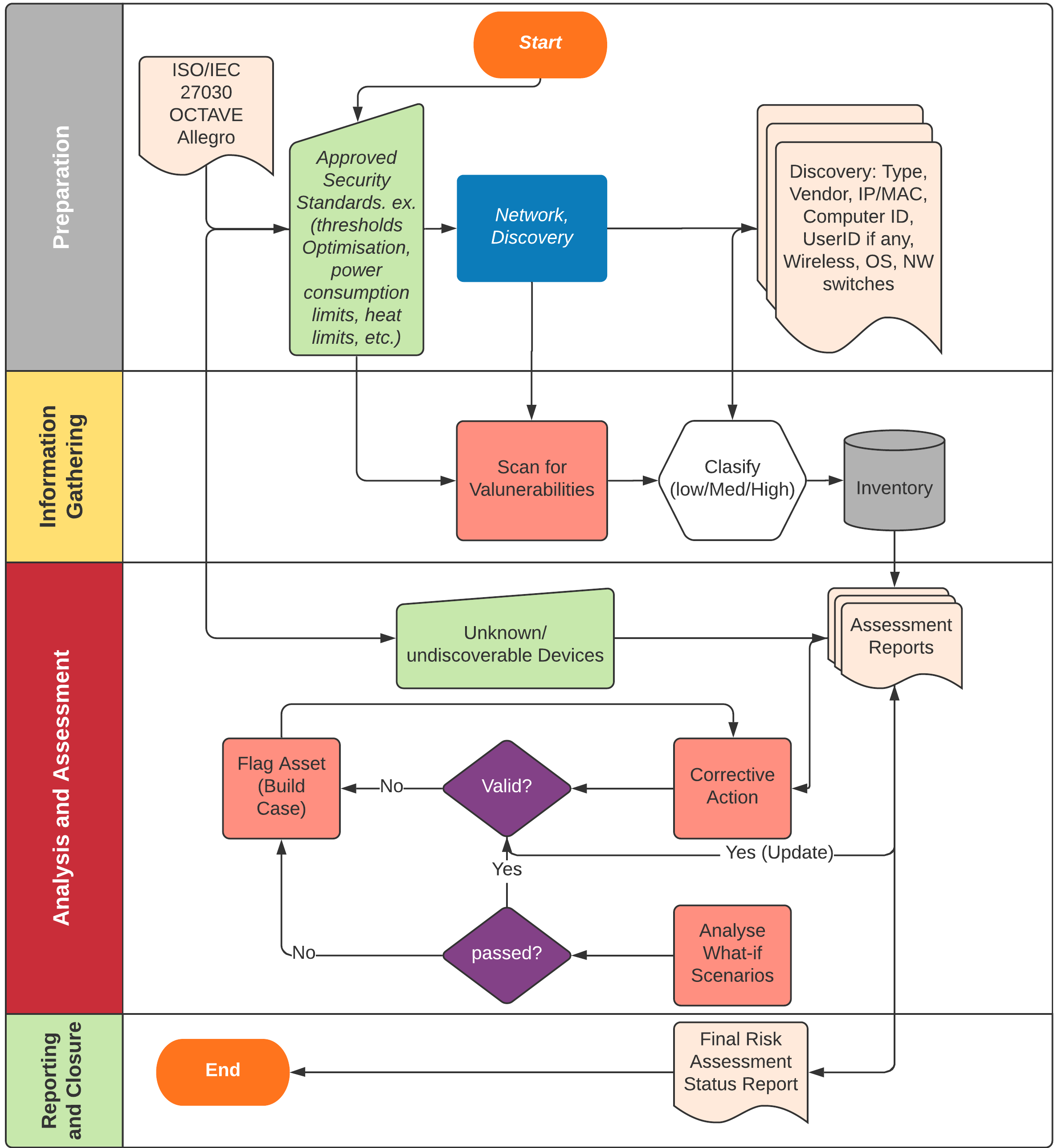}
\caption{IIoT-ARAS System Overview.\cite{b2}}
\label{fig:3}
\end{figure}
\subsection{System Overview}
Given the aforementioned, and concerning IT/OT convergence, a need appears for a customized approach to assess vulnerabilities, threats, and potential risks. Therefore, this research’s contribution builds a risk assessment solution titled IIoT-ARAS that combines and automates a segment of the best practices suggested in the OCTAVE Allegro and ISO/IEC 27030. Current frameworks designed for OT security, such as the Industrial Internet Consortium, ISACA, International Society of Automation, NIST, Technical Support Working Group, etc., all provide necessary details on securing industrial control systems operational technology areas; however, minimum consideration of IT/OT convergence. IIoT-ARAS as shown in figure \ref{fig:3}, on the other hand, is designed to execute automated regular asset inventory checks custom-designed to comply with the IT/OT heterogeneous nature. IIoT-ARAS is designed with four major phases, which are; Preparation, Information Gathering, Analysis and Assessment and Reporting and Closure phases. While these subsystems are highly dependent on each other, however, each performs very unique task as itemized below. 

\begin{itemize}
    \item Preparation: This phase is primarily for the preliminary assessment and asset discovery. Best practices from the ISO/IEC 27030 and OCTAVE Allegro are modeled to extract the approved security standards such as thresholds, optimization, power consumption etc. These standards are used as tuning criteria that are then passed to the network discovery component to identify the network's available devices. The Information discovered will then be passed as input to the next phase. Data such as IP addresses, MAC addresses, IDs, User IDs, wired or wireless networks, Operating Systems, and Network Switches and Routers are organized in a log file as illustrated in Figure \ref{fig:7}. These data are later used in the analysis and assessment phase, e.g. input vector for anomaly detection.
    
    \item Information Gathering - This phase is designed to assist in classifying assets and risk criteria. The discovered assets in phase 1 are scanned for vulnerabilities and the results are classified as low, med or high vulnerabilities based on the modeled security standards as well as the priority and sensitivity of the data. Although, the asset containers principle that comes from OCTAVE Allegro identifies the priority of an asset from the type of information it passes or store, however, in certain scenario, this phase may require some manual effort from network admins to determine and prioritize asset containers based on importance and sensitivity of data. 
    
    \item Analysis and Assessment: Threats on the IIoT network are analyzed and assessed in this phase. The components are responsible for detecting additional unknown or undiscoverable devices and exploring if the detected vulnerabilities are exploitable vulnerabilities that can potentially threaten the system.
    
    \item Reporting and Closure: Finally, the last phase is the reporting and closure. In this phase, the components provide comprehensive and detailed reporting to aid decision-making. By applying the right filtering, we produce detailed log files and illustrative diagrams for a better insight of network status (ex. Figure \ref{fig:7} and \ref {fig:8}). 
\end{itemize}

Furthermore, IIoT-ARAS has an agentless implementation, resulting in a minimal interruption to the OT environment while still utilizing the best security practices to collect data and optimize an acceptable threshold. The tool is under continuous development to support optimization, probability computation, risk evaluations, and contingency plan configuration. 

\section{Implementation}
\subsection{Environment Setup}
OMNeT++ discrete event simulator. OMNeT++ is an object-oriented modular discrete event network simulation framework. It consists of generic architecture. OMNeT++ by itself does not serve as a comprehensive simulator, instead, it provides with tools and structure to write simulations.\cite{b5}
After vigilant study, a number of simulation frameworks are selected to be the base of IIoT-ARAS (Table\ref{tab1}). The heterogeneous environment resulting from IT/OT convergence requires different tools from different domains. 
\subsubsection{Infrastructure}
Due to large number of heterogeneous networks, different frameworks, compatibility issues among several installations, the setup is distributed and being tested over different operating systems environments.
\subsubsection{Limitations}
The setup is intended to perform experiments on simulated networks. IIoT-ARAS supports simulation networks and not tested on actual physical networks due to extensive resources to build ICS/IIoT physical network.

\subsection{Frameworks Implemented}
\begin{figure}[h]
\frame{\includegraphics[width=1\columnwidth] {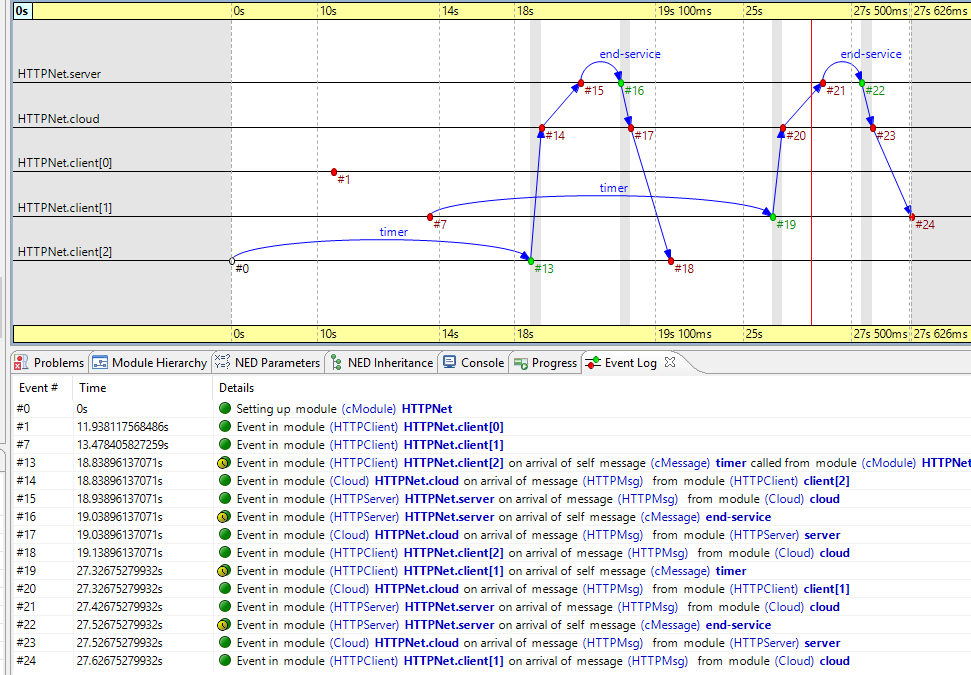}}
\caption{Socket example}
\label{fig:4}
\end{figure}

Omnet++ is a powerful tool that enables simulation frameworks for network connections, communications, attack scenarios, etc. Figure \ref{fig:4} illustrates some of the capabilities of Omnet++ used to connect to physical networks through sockets. To simulate IIoT and assess risks, several existing frameworks as itemized in Table \ref{tab1} are integrated, and new customized tools that provide the needed environment for building and testing IIoT-ARAS are created.

\subsubsection{INET Framework}
INET component is an open-source framework designed for the OMNeT++ simulation-enabling  environment. It provides a set of protocols and other models for research in communication networks. INET is specifically valuable when designing and validating new protocols, or exploring new or unusual scenarios. INET contains most of the TCP stack protocols.
Also, to support many other protocols simulation, several other simulation approaches take INET as a base and extend it into specific routes, such as vehicular networks, overlay/peer-to-peer networks, or LTE..\cite{b6}

\begin{table*}[ht]
\centering
\caption{IIoT-ARAS Implemented Frameworks}
\begin{center}
\begin{tabular}{|c|c|c|c|}
\hline
\cline{1-4}
\textbf{Framework} & \textbf{Omnet++}& \textbf{INET}& \textbf{Purpose} \\
\hline
\cline{1-4}
\hline
\textbf{NETA}&4.2&2.1.0&simulate common attacks in heterogeneous networks\\
\hline
\textbf{Castalia}&5.x&---& networks of low-power embedded and Wireless Sensor Networks (WSN) devices \\
\hline
\textbf{iNetMANET}& 5.x & 4.x& mobile and adhoc routing protocols \\
\hline 
\textbf{ANSA}& 5.0 & 3.0&finding bottlenecks and single-point of failures, configuration errors, faulty network states, etc. \\
\hline 
\textbf{FiCo4OMNET}& 5.0 & 4.0& Supports simulation of fieldbus communication (CAN and FlexRay for ICS) \\
\hline
\textbf{FLoRa}& 5.2.1 & 3.6.3& end-to-end simulations for LoRa networks (IIoT Devices) \\
\hline
\textbf{Fieldbus}& 3.x & --- &performance analysis and evaluation of communication protocols and network configurations of ICS\\
\hline
\cline{1-3}
\end{tabular}
\label{tab1}
\end{center}
\end{table*}
\subsubsection{NETA Framework}
NETwork Attacks (NETA) is an attempt to simulate common attacks in heterogeneous networks using OMNeT++ and the INET-Framework. NETA is aimed to be a practical tool in the network security arena. This tool makes it easy to validate the effectiveness of defense security methods or solutions against network attacks as well as for matching the capabilities of various defense practices.\cite{b6}\cite{b12}
Figure \ref{fig:7} presents and example of a simulated SinkHole Attack.
\begin{figure}[h]
\frame{\includegraphics[width=0.9\columnwidth] {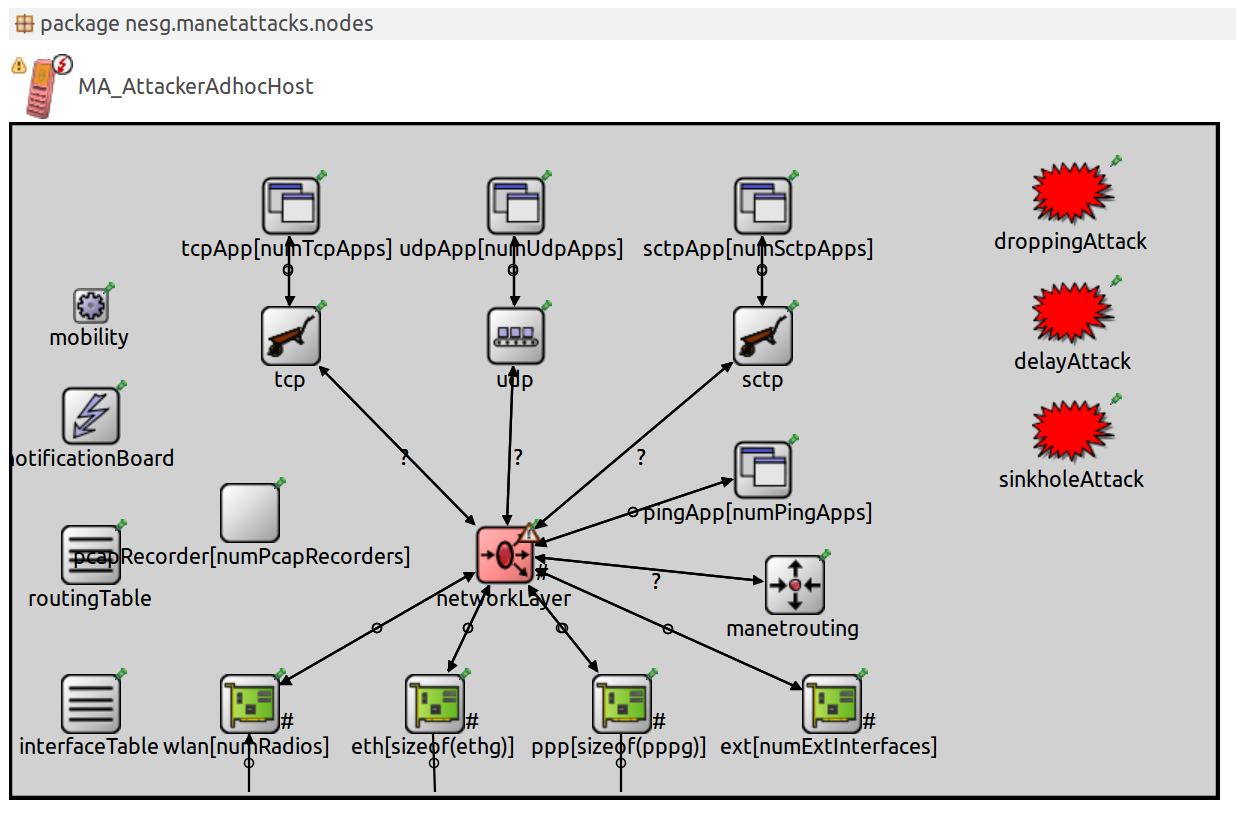}}
\caption{NETA attacks}
\label{fig:8}
\end{figure}

\begin{figure}[h]
\includegraphics[width=1\columnwidth] {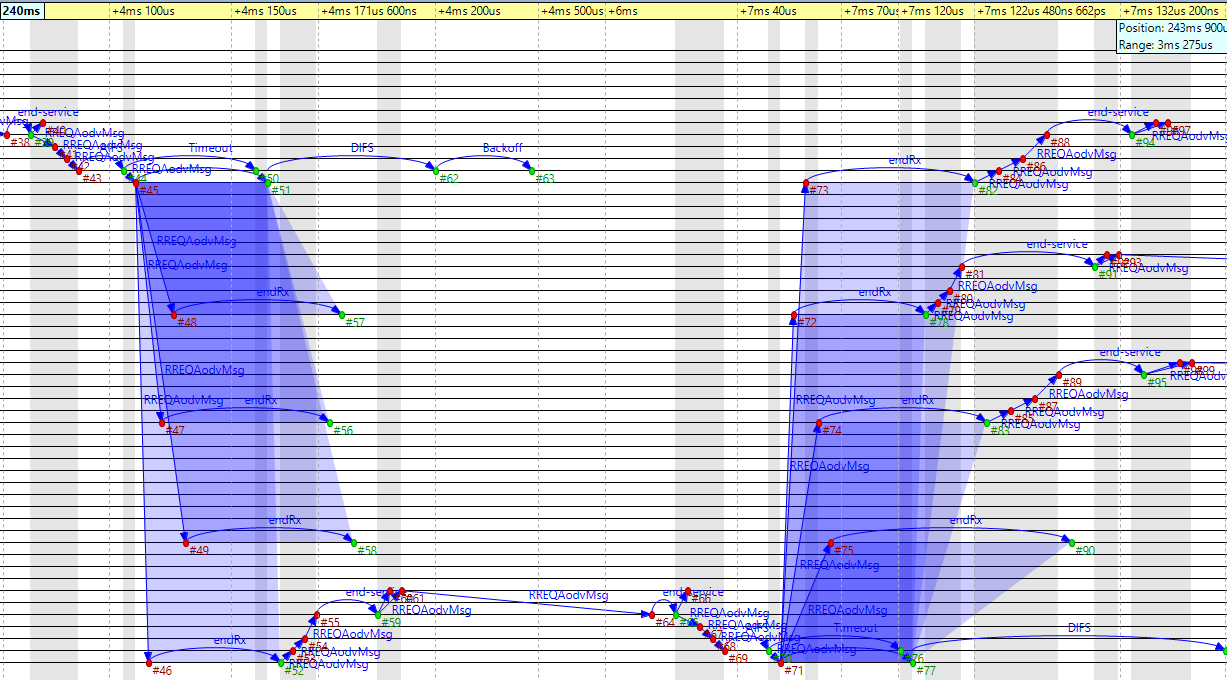}
\caption{SinkHole Attack example}
\label{fig:7}
\end{figure}

\subsubsection{Castalia Framework}
\begin{figure}[h]
\frame{\includegraphics[width=0.9\columnwidth] {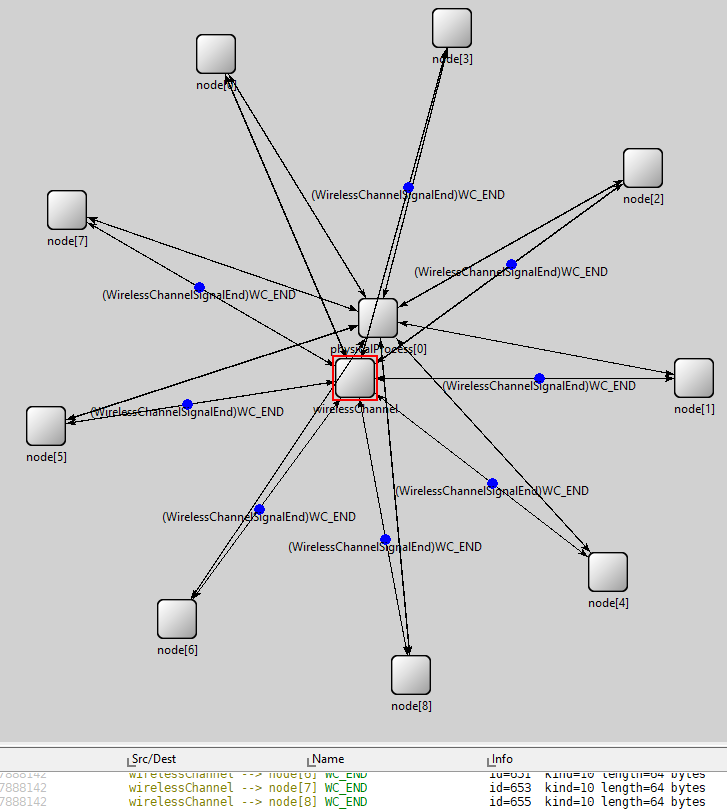}}
\caption{IoT wireless nodes communicates with server}
\label{fig:9}
\end{figure}

Castalia is a simulation tool used to design Wireless Sensor Networks (WSN), Body Area Networks (BAN), and commonly networks of low-power embedded devices as IoT as shown in Figure \ref{fig:9}. Castalia is designed to support researchers to examine distributed algorithms and protocols in like-true wireless channel and radio models, with a realistic node performance especially relating to access to the radio. Castalia’s noticeable features include: the variant exploration of path loss, channel interference and RSSI calculation, physical process modeling, node clock drift, and several popular MAC protocols implemented. Castalia is highly parametric. It provides tools to help run large simulation analyses, develop and present the results graphically.\cite{b6}\cite{b13}

\begin{figure}[h]
\frame{\includegraphics[width=1.0\columnwidth] {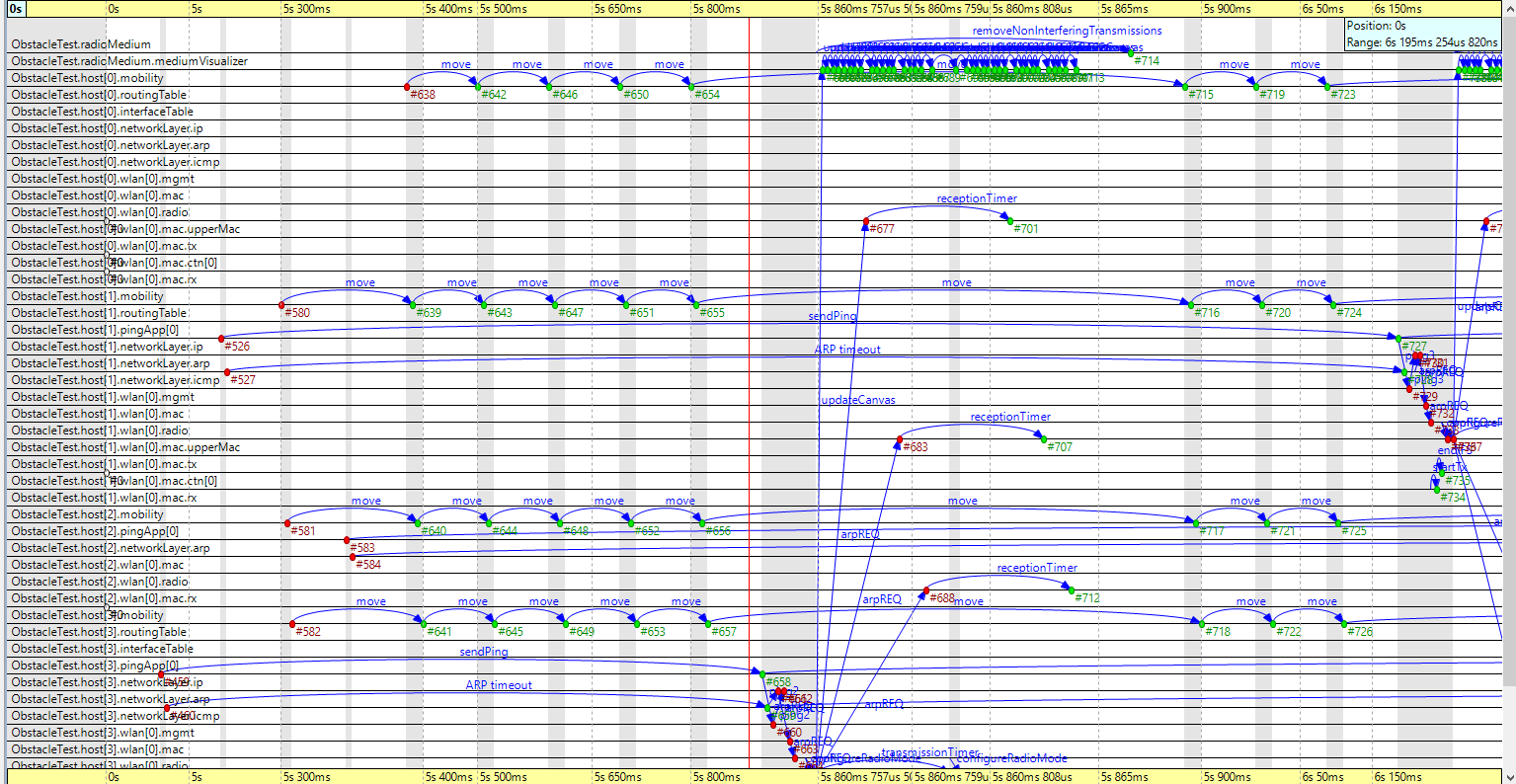}}
\caption{Change in Threshold}
\label{fig:8}
\end{figure}
\subsubsection{INetMANET Framework}
INETMANET 4.x is a branch of the INET Framework 4.x simulator, maintained by Alfonso Ariza Quintana\cite{b6}. INETMANET is kept recent with INET, and extends functionalities with several extra experimental trials and protocols. It is mainly used for mobile ad hoc networks.\cite{b6}\cite{b16}

\subsubsection{ANSA Framework}
The ANSA (Automated Network Simulation and Analysis) project is dedicated to the development of a variety of protocol models, based on RFC specifications and/or reference implementations. The ANSA package extends INET Framework with several protocol models.
ANSA is may be publicly used as the routing/switching baseline for further research initiatives, i.e., in simulations proving (or disproving) certain aspects of networking technologies (e.g., finding bottlenecks and single-point of failures, configuration errors, faulty network states, etc.).
ANSA is a long-term project carried out by researchers and students at Brno University of Technology, Czech Republic.\cite{b6}\cite{b15}

\subsubsection{FiCo4OMNET Framework}
FiCo4OMNeT executes fieldbus communication. Currently, this framework contains two known communication technologies (CAN and FlexRay). Both technologies are applied according to the specification with some adjustments to fit in the simulation platform.
Implemented by the CoRE (Communication over Realtime Ethernet) research group with support from the INET (Internet Techologies) research group at the HAW Hamburg (Hamburg University of Applied Sciences).\cite{b7}


\subsubsection{FLoRa Framework}
FLoRa is a specialized Framework for LoRa. It is a simulation platform for handling
end-to-end simulations for LoRa networks. It is based on the OMNeT++ network simulator
and exploits components from the INET framework. FLoRa utilizes OMNeT 5.x and INET 3.x. 
FLoRa is authored by Mariusz Slabicki and Gopika Premsankar.\cite{b6}\cite{b14}
FLoRa basically, allows the creation of LoRa networks with specific modules designed
for LoRa nodes, gateways and network servers. Application logic can be arranged as
standalone modules that connect to the network server. The network server and connected
nodes support dynamic management of configuration parameters through Adaptive Data Rate
\begin{figure}[h]
\includegraphics[width=0.9\columnwidth,trim=0 0 0 0,clip]{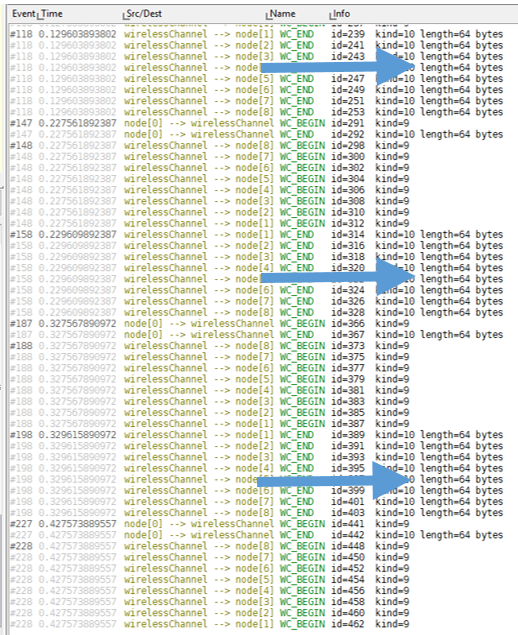}
\caption{Log files for predicting anomalies}
\label{fig:7}
\end{figure}
(ADR). Furthermore, the energy consumption statistics are collected and calculated in
all nodes.\cite{b6}

\subsection{IIoT-ARAS Phases Simulation Frameworks} 
IIoT-ARAS consists of 4 main phases presenting the lifecycle of the risk assessment tool.
The steps are customized and adopted from OCTAVE Allegro and ISO/IEC information security frameworks.

\subsubsection{Preparation Phase}
INET is a core component that provides the ability to build and simulate several types of networks. INETMANET comes as an extension to INET to allow simulations of mobile and ad-hoc networks. Using these two frameworks, we developed a multiple heterogeneous simulated networks depicting multiple scenario. The IT/OT networks are a mix of wired, wireless, low power, and embedded devices communicating together in a single system.

Before we begin the network discovery in the preparation, we first configure the network architecture. This is given in a network description file (.ned) in Omnet++. The file contains the topology of the network as well as the modeled ISO/OCTAVE allegro security standards. Subsequently, an initialization file (.ini) is created to control the parameters and execution of the network description file. Furthermore, we integrate a customized component in the .ini file to perform ping sweep. 

At this phase the simulated network is ready to perform network discovery using the given parameters in the (.ned) file. These parameters can further be fine-tuned using an optimized component written in c++.


\subsubsection{Information Gathering Phase}
In this phase, a combination of frameworks are used for a much deeper stress testing and vulnerability scan of the discovered assets. For instance, ANSA was used for enabling further configuration of active devices to predict network behavior. ANSA contains many improvements to networking protocols used in INET framework. Supported protocols include Gateway Load-Balancing Protocol (GLBP), Intermediate System to Intermediate System (IS-IS), and many layer 2 (L2) management protocols. This allows gathering further information about routing and switching\cite{b6}\cite{b15}. FLoRa framework is used in this phase to support end-to-end simulations for gateways and low power devices like IIoT. Also, FLoRa supports gathering information about power consumption which is considered valuable in risk analysis\cite{b14}. To have the ability to simulate Wireless Sensor Networks (WSN), Body Area Networks (BAN), and embedded devices, Castalia framework is implemented in this phase to provide realistic information about path loss, interference, and insight about MAC protocols in use. In the Information Gathering phase, a support for Industrial Control Systems is required\cite{b13}. FiCo4OMNET framework is designed to give support for CAN and Flexray technologies to enable simulation of remote mobility\cite{b7}. Finally, Fieldbus framework is utilized due to its capabilities of simulating ICS networks and collecting information\cite{b6}.


\subsubsection{Analysis and Assessment Phase}
In IIoT-ARAS implementation, the component in this phase will examine the network strength and availability by examining the network bandwidth, bottlenecks, points of failure and network configuration errors in general using ANSA framework . In addition, simulated attacks is performed to test compliance with CIA triad and to examine if the vulnerability will result in threat using the NETA framework.

 

\subsubsection{Reporting and Closure Phase}

In this phase, the use of all frameworks is necessary to guarantee detailed reporting and closure. The phase is under development to calculate probability and impacts, threat and risk predictions, in addition to vulnerable configurations and setups. Each of the previously used framework (INET, INETMANET, ANSA, NETA, FLoRa, Castalia, FiCo4OMNET, Fieldbus) provides a number of logs and data that can be used to present status of the network. Integrating and analyzing collected data will aid in successful risk assessment and prediction. 

 

\section{IIoT-ARAS Initial Testing}

\subsection{IP Dropping Attack}

\begin{figure}[h]
\frame{\includegraphics[width=1.0\columnwidth,trim=0 0 0 0,clip]{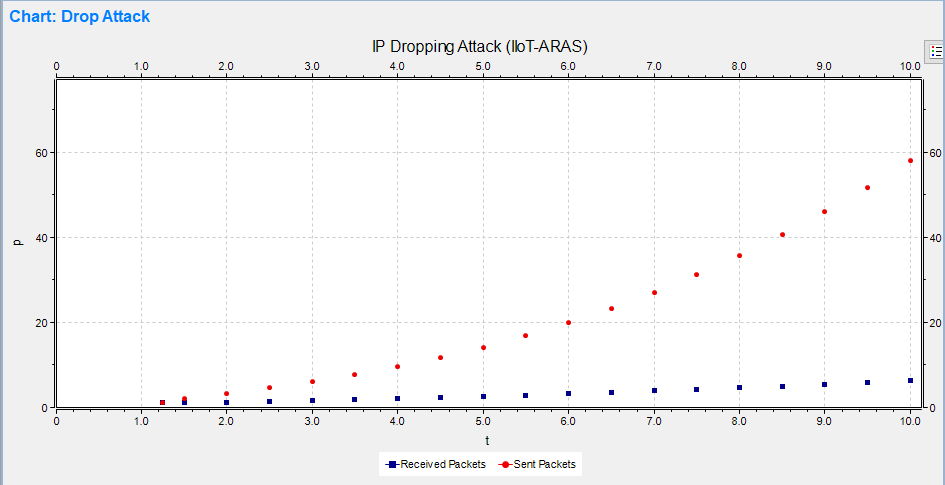}}
\caption{Sent vs. Received Packets}
\label{fig:8}
\end{figure}
In IP Dropping Attack, nodes affected will drop received packets instead of forwarding them to intended party. The attack can compromise bandwidth, quality of service (QoS), and availability of network resources. Figure \ref{fig:8} shows the ratio between sent and received packets. It is clear that the packet loss ratio is significant. For evaluation purposes, the simulation of attack succeeded and expected results are generated. For future testing and evaluation, we will add more testing and performance metrics like Packet Delivery Ratio (PDR) and the Dropping Ratio (DR):
\begin{equation}
    Packets Delivered (Pd)/Packets Transmitted (Pt) 
\end{equation}
\begin{equation}
    Packets Lost (Pl)/Packets Transmitted (Pt) 
\end{equation}
Pd: Packets Delivered, Pt: Packets transmitted, Pl: Packets lost.
\\

\section{Limitation and Future Work}
The main challenge we face is to test and assess a heterogeneous network that was built using different components that communicates by several protocols. The ever changing nature and incompatibility issues in the IT/OT convergence world will never end. For testing purposes, we ran and analyze sample attacks on simulated network to ensure validity. But nonetheless, we will explore the performance of IIoT-ARAS on much larger scale real network. In Addition, we will work on integrating all different components of IIoT-ARAS into one structured system, introduce socket programming to be able to connect the tool to real physical networks, create more simulated network attacks for testing, and work on enhancing assets discovery by adopting enhanced version of known protocols like uPNP, SNMP, SolarWind's Ping Sweep and others. Furthermore, we will work on creating a simulation framework for BACnet protocol being one of the most common ICS used protocols\cite{b17}. As a next phase, we consider introducing Machine Learning principles to IIoT-ARAS to aid self-decision and protection by utilizing in-memory objects.

\section{Conclusion} 
The IIoT-ARAS is a tool used to evaluate and assess vulnerabilities and risks associated based on OCTAVE Allegro and initial ISO/IEC 27030 guidelines. The main purpose is to examine the IT/OT convergence where information technology meets industrial operations. A basic IT/OT network is evaluated against a number of known attacks to record the change in threshold and behavior. This will present an input to further study of malware detection and prevention. The work is a continuous effort and will be tested against physical IT/OT networks.

\section*{Acknowledgement}
This work is supported by the National Science Foundation (NSF) under Grant Number 1850054.

\end{document}